\documentclass[12pt]{article}

\usepackage{amsmath,amssymb}
\usepackage{amsmath,amscd}
\usepackage{bm}
\usepackage{mathrsfs}
\usepackage{latexsym}
\usepackage{multirow}
\usepackage{float}
\usepackage[usenames]{color}
\usepackage{indentfirst}  
\usepackage{cite}
\usepackage[dvips]{graphicx}
\usepackage{color}

\allowdisplaybreaks[4]

\newcommand{{\Slashp}}{p\!\!\!\!\!\big/}
\newcommand{{\Slashq}}{q\!\!\!\!\!\big/}

\makeatletter
\@addtoreset{equation}{section}

\makeatother

\begin{document}

\begin{titlepage}
\thispagestyle{empty}
~~\\
\vspace{5mm} 

\begin{center}
{\LARGE Left-right symmetry, orbifold $S^1/Z_2$,\\
and radiative breaking of $U(1)_{\rm R} \times U(1)_{\rm B-L}$}
\end{center}

\begin{center}
\lineskip .45em
\vskip1.5cm
{\large Yugo Abe$^a$\footnote{E-mail: yugoabe@miyakonojo.kosen-ac.jp}, 
Yuhei Goto$^b$\footnote{E-mail: y-goto@keio.jp}
and
Yoshiharu Kawamura$^c$\footnote{E-mail: haru@azusa.shinshu-u.ac.jp}}

\vskip 1.5em
${}^a\,${\large\itshape 
National Institute of Technology, Miyakonojo College, 
Miyakonojo 885-8567, Japan}\\[1mm]
${}^b\,${\large\itshape 
Research and Education Center for Natural Science, 
Keio University, Yokohama 223-8521, Japan}\\[1mm]
${}^c\,${\large\itshape Department of Physics, Shinshu University, 
Matsumoto 390-8621, Japan}

 \vskip 4.5em
\end{center}



\begin{abstract}
We study the origin of electroweak symmetry
under the assumption that 
$SU(4)_{\rm C} \times SU(2)_{\rm L} \times SU(2)_{\rm R}$
is realized on a five-dimensional space-time.
The Pati-Salam type gauge symmetry is reduced to
$SU(3)_{\rm C} \times SU(2)_{\rm L} \times U(1)_{\rm R}
\times U(1)_{\rm B-L}$ by orbifold breaking mechanism on
the orbifold $S^1/Z_2$.
The breakdown of residual gauge symmetries
occurs radiatively via the Coleman-Weinberg mechanism,
such that the $U(1)_{\rm R} \times U(1)_{\rm B-L}$ symmetry
is broken down to $U(1)_{\rm Y}$ by the vacuum expectation value
of an $SU(2)_{\rm L}$ singlet scalar field
and the $SU(2)_{\rm L} \times U(1)_{\rm Y}$ symmetry is broken down to
the electric one $U(1)_{\rm EM}$ by the vacuum expectation value
of an $SU(2)_{\rm L}$ doublet scalar field regarded as the Higgs doublet.
The negative Higgs squared mass term
is originated from an interaction between the Higgs doublet 
and an $SU(2)_{\rm L}$ singlet scalar field as a Higgs portal.
The vacuum stability is recovered
due to the contributions from Kaluza-Klein modes of gauge bosons.
\end{abstract}

\end{titlepage}

\section{Introduction}

The discovery of the Higgs boson~\cite{ATLAS,CMS},
the last piece of the standard model (SM) particles,
kicks off a new stage of physics beyond the SM.
Mysteries concerning the Higgs boson have thickened
because any evidences 
from new physics such as supersymmetry 
and compositeness have not been discovered.

One of big mysteries is what the origin of electroweak scale is
or how the vacuum expectation value (VEV) 
of the Higgs boson, $v = 246$ GeV, is understood.
To unveil the riddle, we need to uncover the origin 
of Higgs potential, in particular, a mass term therein.
Another one is why the vacuum is stable enough
after the breakdown of electroweak symmetry.
With the Higgs quartic coupling constant 
$\lambda \doteqdot 0.129$ estimated from
the observed Higgs mass $m_h \doteqdot 125.1$ GeV, 
we encounter the vacuum stability problem
that $\lambda$ becomes negative 
at around $10^{7}$ GeV
and the vacuum can decay.

In this paper, we tackle these problems 
through the extensions of gauge symmetries and space-time.
Concepts such as simplicity and variety are also adopted
on a case-by-case basis.
The SM gauge symmetry can be extended to contain
a left-right symmetry.
A typical one is the gauge group
$G_{\rm PS} \equiv SU(4)_{\rm C} \times SU(2)_{\rm L} \times SU(2)_{\rm R}$
in the Pati-Salam model~\cite{P&S}.
The space-time can be expanded to include extra dimensions.
The orbifold $S^1/Z_2$ is used as an extra space,
because it is simple and has several advantages.
Different breaking mechanisms are utilized 
for the breakdown of gauge symmetry $G_{\rm PS}$
into $SU(3)_{\rm C} \times U(1)_{\rm EM}$,
presuming that nature respects diversity. 

We give an outline of our model.
Particle physics above some high-energy scale $M_{\rm PS}$
is described by a gauge theory with $G_{\rm PS}$
on the five-dimensional (5D) space-time including
$S^1/Z_2$ as an extra dimension.
The gauge symmetry $G_{\rm PS}$ is reduced to
$G_{3211} \equiv SU(3)_{\rm C} \times SU(2)_{\rm L} \times U(1)_{\rm R}
\times U(1)_{\rm B-L}$ by orbifold breaking mechanism\footnote{
The orbifold breaking mechanism was originally proposed 
in superstring theory~\cite{DHV&W1,DHV&W2}.
The $Z_2$ orbifolding 
was used in superstring theory~\cite{A} 
and heterotic M-theory~\cite{H&W1,H&W2}.
In field theoretical models, it was applied to 
the reduction of global supersymmetry~\cite{M&Peskin,P&Q}, which is
an orbifold version of Scherk-Schwarz mechanism~\cite{S&S,S&S2}, 
and then to the reduction of gauge symmetry~\cite{K1,K2}.
The left-right symmetric models on 5D space-time were proposed 
in \cite{M&N,M&P}, and phenomenologies on 
gauge bosons and matter fields were studied intensively
based on the gauge group $SU(2)_{\rm L} \times SU(2)_{\rm R}
\times U(1)_{\rm B-L}$.}.
The breakdown of residual gauge symmetries
occurs radiatively via the Coleman-Weinberg mechanism\footnote{
The Coleman-Weinberg mechanism was originally proposed 
by S.~Coleman and E.~Weinberg~\cite{C&W},
and used in left-right symmetric models~\cite{S&M,C,HL&S,BDM&Z}
and a minimal extension of the SM 
with a SM singlet and an extra $U(1)$ symmetry~\cite{Hempfling}.}.
In concrete, the $U(1)_{\rm R} \times U(1)_{\rm B-L}$ symmetry
is broken down to $U(1)_{\rm Y}$ 
by the VEV $v_{\rm R}$
of an $SU(2)_{\rm L}$ singlet scalar field.
Then, a gauge boson corresponding to the broken $U(1)$ symmetry
acquires a mass $M_{Z_{\rm LR}}$ of $O(v_{\rm R})$.
The $SU(2)_{\rm L} \times U(1)_{\rm Y}$ symmetry is broken down to
the electric one $U(1)_{\rm EM}$ by the VEV
of an $SU(2)_{\rm L}$ doublet scalar field regarded as the Higgs doublet.
If the $SU(2)_{\rm L}$ singlet scalar field is replaced by its VEV,
we obtain the Higgs potential including
a negative squared mass term
originated from an interaction between the Higgs doublet 
and an $SU(2)_{\rm L}$ singlet scalar field as a Higgs portal.
The vacuum stability is recovered 
due to the contributions from Kaluza-Klein modes 
of gauge bosons appearing at a compactification scale $M_{\rm c}$.

This paper is organized as follows.
In the next section, we formulate a 5D Pati-Salam model.
We examine the Coleman-Weinberg mechanism
and the vacuum stability
in a four-dimensional (4D) model with $G_{3211}$
in Sect.~3, 
In the last section, we give conclusions and discussions.

\section{Five-dimensional Pati-Salam model}
\label{PSmodel}

The space-time is assumed to be factorized into 
a product of 4D Minkowski space-time $M^4$ 
and the orbifold $S^1/Z_2$, whose coordinates are denoted by 
$x^\mu$ (or $x$) ($\mu = 0, 1, 2, 3$) and $y$, respectively.
The 5D notation $x^M$ ($M = 0, 1, 2, 3, 5$) is also used
with $x^5 = y$.
The $S^1/Z_2$ is obtained 
by dividing the circle $S^1$ (with the identification $y\sim y+2\pi R$)
by the $Z_2$ transformation $y \rightarrow -y$.
Then, the point $y$ is identified with $-y$ on $S^1/Z_2$,
and the space is regarded as an interval with length $\pi R$
($R$ being the radius of $S^1$).

In the following, we formulate a Pati-Salam model 
on $M^4 \times S^1/Z_2$.
First we present particle contents in Table \ref{T-PS}.
In most cases, we pay attention to bosons
under the assumption that matter fields (quarks and leptons)
live on the 4D brane at $y=0$.
\begin{table}[htbp]
\caption{Gauge quantum numbers of bosons in 5D Pati-Salam model.}
\label{T-PS}
\begin{center}
\begin{tabular}{c|c|c|c}
\hline                              
bosons & $SU(4)_{\rm C}$ & $SU(2)_{\rm L}$ & $SU(2)_{\rm R}$ 
\\ \hline 
$G_M(x, y)$ & $\bm{15}$ & $\bm{1}$ & $\bm{1}$ \\
$W_{{\rm L}M}(x, y)$ & $\bm{1}$ & $\bm{3}$ & $\bm{1}$ \\
$W_{{\rm R}M}(x, y)$ & $\bm{1}$ & $\bm{1}$ & $\bm{3}$ \\
$\varPhi_{\rm L}(x, y)$ & ${\bm{4}}$ & $\bm{2}$ & $\bm{1}$ \\
$\varPhi_{\rm R}(x, y)$ & $\overline{\bm{4}}$ & $\bm{1}$ & $\bm{2}$ \\
$\varPhi_{\rm B}(x, y)$ & $\bm{1}$ & $\bm{2}$ & $\bm{2}$ \\
\hline
\end{tabular}
\end{center}
\end{table}
The gauge bosons possess several components such that
\begin{eqnarray}
&~& G_M(x, y) = \sum_{a=1}^{15} G_M^a(x, y) T_{\rm C}^a,~~
\nonumber \\
&~& W_{{\rm L}M}(x, y) = \sum_{a=1}^{3} W_{{\rm L}M}^a(x, y) T_{\rm L}^a,~~
W_{{\rm R}M}(x, y) = \sum_{a=1}^{3} W_{{\rm R}M}^a(x, y) T_{\rm R}^a,
\label{gauge-5D}
\end{eqnarray}
where $T_{\rm C}^a$, $T_{\rm L}^a$ and $T_{\rm R}^a$
are generators of $SU(4)_{\rm C}$, $SU(2)_{\rm L}$ 
and $SU(2)_{\rm R}$, respectively.
We need a scalar field
$\varPhi_{\rm B}(x, y)$ that obeys the bi-fundamental
representation under $SU(2)_{\rm L} \times SU(2)_{\rm R}$,
to construct Yukawa interactions on the brane.
The Lagrangian density for bosons is given by
\begin{eqnarray}
\hspace{-1.2cm}
&~& \mathscr{L}_{\rm 5D}
= -\frac{1}{4} \sum_{a=1}^{15} G_{MN}^a G^{a MN}
-\frac{1}{4} \sum_{a=1}^{3} W_{{\rm L}MN}^a W_{\rm L}^{a MN}
-\frac{1}{4} \sum_{a=1}^{3} W_{{\rm R}MN}^a W_{\rm R}^{a MN}
\nonumber \\
\hspace{-1.2cm}
&~&~~~~~~~~~ + (D_M \varPhi_{\rm L})^{\dagger} (D^M \varPhi_{\rm L}) 
+ (D_M \varPhi_{\rm R})^{\dagger} (D^M \varPhi_{\rm R}) 
+ {\rm tr}(D_M \varPhi_{\rm B})^{\dagger} (D^M \varPhi_{\rm B}) - V_{\rm 5D},
\label{L-5D}
\end{eqnarray}
where $G_{MN}^a$, $W_{{\rm L}MN}^a$ and $W_{{\rm R}MN}^a$
are field strengths of $SU(4)_{\rm C}$, $SU(2)_{\rm L}$ 
and $SU(2)_{\rm R}$ gauge bosons, respectively.
The covariant derivative $D_{M}$ 
and the scalar potential $V_{\rm 5D}$
are given by
\begin{eqnarray}
&~& D_M = \partial_M + i g_4 \sum_{a=1}^{15} G_M^a T_{\rm C}^a
+ i g_{\rm L} \sum_{a=1}^{3} W_{{\rm L}M}^a T_{\rm L}^a
+ i g_{\rm R} \sum_{a=1}^{3} W_{{\rm R}M}^a T_{\rm R}^a,
\label{D_M}\\
&~& V_{\rm 5D}=\lambda_{\rm L} |\varPhi_{\rm L}|^4
+\lambda_{\rm R} |\varPhi_{\rm R}|^4 
+ \lambda_{\rm B1} {\rm tr}\left(|\varPhi_{\rm B}|^2 |\varPhi_{\rm B}|^2\right)
+ \lambda_{\rm B2} \left({\rm tr}|\varPhi_{\rm B}|^2\right)^2
\nonumber \\
&~& ~~~~~ + \lambda_{\rm LR} |\varPhi_{\rm L}|^2  |\varPhi_{\rm R}|^2
+ \lambda_{\rm LB} |\varPhi_{\rm L}|^2  {\rm tr}|\varPhi_{\rm B}|^2
+ \lambda_{\rm RB} |\varPhi_{\rm R}|^2  {\rm tr}|\varPhi_{\rm B}|^2,
\label{V-5D}
\end{eqnarray}
respectively.
If we require the left-right symmetry
that the theory should be invariant under
the exchange $(W_{{\rm L}M}^a, \varPhi_{\rm L})$ 
into $(W_{{\rm R}M}^a, \varPhi_{\rm R})$,
we obtain the conditions among couplings:
\begin{eqnarray}
g_{\rm L} = g_{\rm R},~~ \lambda_{\rm L} = \lambda_{\rm R},~~
\lambda_{\rm LB} = \lambda_{\rm RB}.
\label{cond-5D}
\end{eqnarray}
We suppose that all scalar fields have no bulk masses.

From the requirement that 
the Lagrangian density should be invariant under the translation
$T:y \to y + 2\pi R$ and
the $Z_2$ transformation $P_0:y \to -y$
or it should be a single-valued function on the 5D space-time,
non-trivial boundary conditions (BCs) of fields
are allowed on $S^1/Z_2$.

We impose the following BCs on $G_M$,
\begin{eqnarray}
\hspace{-0.6cm}
&~& G_{\mu}(x, -y)=G_{\mu}(x, y),~~ G_{5}(x, -y)=-G_{5}(x, y),
\label{GM-BC1}\\
\hspace{-0.6cm}
&~& G_\mu(x, 2\pi R-y)=U_{\rm C} G_\mu(x, y) U_{\rm C}^{-1},~~
G_5(x, 2\pi R-y)=-U_{\rm C} G_5(x, y) U_{\rm C}^{-1},
\label{GM-BC2}
\end{eqnarray}
where $U_{\rm C} = {\rm diag}(1, 1, 1, -1)$.
We use the $Z_2$ transformation $P_1:y \to 2\pi R -y$
in place of $T:y \to y + 2\pi R$.
Then, $G_M$ are given by the Fourier expansions:
\begin{eqnarray}
\hspace{-0.6cm}&~& G_\mu^{a}(x, y)=\frac{1}{\sqrt{2\pi{R}}}G_{\mu}^{(0)a}(x)
+\frac{1}{\sqrt{\pi{R}}}\sum_{n=1}^{\infty}G_{\mu}^{(n)a}(x)\cos\frac{ny}{R}
~~~ (a=1, \cdots, 8, 15),
\label{Gmu++}\\
\hspace{-0.6cm}&~& G_\mu^{a}(x, y)
=\frac{1}{\sqrt{\pi{R}}}\sum_{n=1}^{\infty}G_{\mu}^{(n)a}(x)
\cos\frac{\left(n-\frac{1}{2}\right)y}{R}
~~~ (a=9, \cdots, 14),
\label{Gmu+-}\\
\hspace{-0.6cm}&~& G_5^a(x, y)=\frac{1}{\sqrt{\pi{R}}}\sum_{n=1}^{\infty}
G_{5}^{(n)a}(x)\sin\frac{ny}{R} ~~~ (a=1, \cdots, 8, 15),
\label{G5--}\\
\hspace{-0.6cm}&~& G_5^{a}(x, y)
=\frac{1}{\sqrt{\pi{R}}}\sum_{n=1}^{\infty}G_{5}^{(n)a}(x)
\sin\frac{\left(n-\frac{1}{2}\right)y}{R}
~~~ (a=9, \cdots, 14).
\label{G5-+}
\end{eqnarray}
Only $G_{\mu}^a$ ($a=1, \cdots, 8, 15$) have
$y$-independent modes with $n=0$ called zero modes,
and $G_{\mu}^{(0)a}(x)$ ($a=1, \cdots, 8$) and 
$G_{\mu}^{(0)15}(x)$ are identified as the 4D gluons
and the 4D $U(1)_{\rm B-L}$ gauge boson, respectively.
We denote them as $G_{\mu}^a(x)$ and $N_{\mu}(x)$, respectively.

We impose the following BCs on $W_{{\rm L}M}$,
\begin{eqnarray}
&~& W_{{\rm L}\mu}(x, -y)=W_{{\rm L}\mu}(x, y),~~ 
W_{{\rm L}5}(x, -y)=-W_{{\rm L}5}(x, y),
\label{WLM-BC1}\\
&~& W_{{\rm L}\mu}(x, 2\pi R-y)
=W_{{\rm L}\mu}(x, y),~~
W_{{\rm L}5}(x, 2\pi R-y)=-W_{{\rm L}5}(x, y)
\label{WLM-BC2}
\end{eqnarray}
and then we obtain the zero modes $W_{{\rm L}\mu}^{(0)a}(x)$
($a=1, 2, 3$) identified as 
the 4D $SU(2)_{\rm L}$ weak bosons
and denote them as $W_{\mu}^{a}(x)$.

We impose the following BCs on $W_{{\rm R}M}$,
\begin{eqnarray}
\hspace{-1.7cm}&~& W_{{\rm R}\mu}(x, -y)=W_{{\rm R}\mu}(x, y),~~ 
W_{{\rm R}5}(x, -y)=-W_{{\rm R}5}(x, y),
\label{WRM-BC1}\\
\hspace{-1.7cm}&~& W_{{\rm R}\mu}(x, 2\pi R-y)
=U_{\rm R} W_{{\rm R}\mu}(x, y) U_{\rm R}^{-1},~~
W_{{\rm R}5}(x, 2\pi R-y)=-U_{\rm R} W_{{\rm R}5}(x, y) U_{\rm R}^{-1},
\label{WRM-BC2}
\end{eqnarray}
where $U_{\rm R} = {\rm diag}(1, -1)$.
Then, we obtain the zero modes $W_{{\rm R}\mu}^{(0)3}(x)$
regarded as a $U(1)$ gauge boson.
We denote $W_{{\rm R}\mu}^{(0)3}(x)$
and its $U(1)$ gauge group as $R_{\mu}(x)$ and $U(1)_{\rm R}$, respectively.

For scalar fields, the following BCs are imposed on,
\begin{eqnarray}
&~& \varPhi_{\rm L}(x, -y)=-\varPhi_{\rm L}(x, y),~~
\varPhi_{\rm L}(x, 2\pi R-y)=-U_{\rm C}\varPhi_{\rm L}(x, y),
\label{PhiL-BC}\\
&~&  \varPhi_{\rm R}(x, -y)=-U_{\rm R} \varPhi_{\rm R}(x, y),~~
\varPhi_{\rm R}(x, 2\pi R-y)=-U_{\rm C} \varPhi_{\rm R}(x, y),
\label{PhiR-BC}\\
&~&  \varPhi_{\rm B}(x, -y)=\varPhi_{\rm B}(x, y),~~
\varPhi_{\rm B}(x, 2\pi R-y)=U_{\rm R} \varPhi_{\rm B}(x, y).
\label{PhiB-BC}
\end{eqnarray}
Then, zero modes appear from the lower component
of $\varPhi_{\rm R}$ 
and the upper component of $\varPhi_{\rm B}$
concerning $SU(2)_{\rm R}$,
and they are denoted as $\phi_{\rm R}(x)$ and $\phi(x)$, respectively.
Here, $\phi_{\rm R}(x)$ is the $SU(2)_{\rm L}$ singlet scalar field
and $\phi(x)$ is the $SU(2)_{\rm L}$ doublet scalar field.
The $\phi(x)$ is regarded as the Higgs doublet in the SM.

We list gauge quantum numbers and mass spectra 
of bosons after compactification in Table \ref{T-PSsp}.
\begingroup
\renewcommand{\arraystretch}{1.2}
\begin{table}[htbp]
\caption{Gauge quantum numbers of bosons 
after compactification in 5D Pati-Salam model.}
\label{T-PSsp}
\begin{center}
\begin{tabular}{c|c|c|c|c|c|c}
\hline                              
bosons & $SU(3)_{\rm C}$ & $SU(2)_{\rm L}$ & $Q_{\rm R}$ 
& $Q_{\rm B-L}$ & $(P_0, P_1)$ & mass \\
\hline\hline
$G_{\mu}^{(n)a}(x, y)$ $(a=1 \sim 8)$
& $\bm{8}$ & $\bm{1}$ & $0$ 
& $0$ & $(+1, +1)$ & $\frac{n}{R}$ \\
$G_{\mu}^{(n)a}(x, y)$ $(a=9 \sim 14)$ & $\bm{3}$ & $\bm{1}$ & $0$ 
& $\frac{2}{3}$ & $(+1, -1)$ & $\frac{n-\frac{1}{2}}{R}$ \\
& $\overline{\bm{3}}$ & $\bm{1}$ & $0$ 
& $-\frac{2}{3}$ & $(+1, -1)$ & $\frac{n-\frac{1}{2}}{R}$ \\
$G_{\mu}^{(n)15}(x, y)$ & $\bm{1}$ & $\bm{1}$ & $0$ 
& $0$ & $(+1, +1)$ & $\frac{n}{R}$ \\
\hline
$W_{{\rm L}\mu}^{(n)a}(x, y)$ $(a=1, 2, 3)$ & $\bm{1}$ & $\bm{3}$ & 0 
& $0$ & $(+1, +1)$ & $\frac{n}{R}$ \\
\hline
$W_{{\rm R}\mu}^{(n)3}(x, y)$ & $\bm{1}$ & $\bm{1}$ & $0$ 
& $0$ & $(+1, +1)$ & $\frac{n}{R}$ \\
$W_{{\rm R}\mu}^{(n)+}(x, y)$ & $\bm{1}$ & $\bm{1}$ & $1$ 
& $0$ & $(+1, -1)$ & $\frac{n-\frac{1}{2}}{R}$ \\
$W_{{\rm R}\mu}^{(n)-}(x, y)$ & $\bm{1}$ & $\bm{1}$ & $-1$ 
& $0$ & $(+1, -1)$ & $\frac{n-\frac{1}{2}}{R}$ \\
\hline
$G_{5}^{(n)a}(x, y)$ $(a=1 \sim 8)$
& $\bm{8}$ & $\bm{1}$ & $0$ 
& $0$ & $(-1, -1)$ & $\frac{n}{R}$ \\
$G_{5}^{(n)a}(x, y)$ $(a=9 \sim 14)$ & $\bm{3}$ & $\bm{1}$ & $0$ 
& $\frac{2}{3}$ & $(-1, +1)$ & $\frac{n-\frac{1}{2}}{R}$ \\
& $\overline{\bm{3}}$ & $\bm{1}$ & $0$ 
& $-\frac{2}{3}$ & $(-1, +1)$ & $\frac{n-\frac{1}{2}}{R}$ \\
$G_{5}^{(n)15}(x, y)$ & $\bm{1}$ & $\bm{1}$ & $0$ 
& $0$ & $(-1, -1)$ & $\frac{n}{R}$ \\
\hline
$W_{{\rm L}5}^{(n)a}(x, y)$ $(a=1, 2, 3)$ & $\bm{1}$ & $\bm{3}$ & 0 
& $0$ & $(-1, -1)$ & $\frac{n}{R}$ \\
\hline
$W_{{\rm R}5}^{(n)3}(x, y)$ & $\bm{1}$ & $\bm{1}$ & $0$ 
& $0$ & $(-1, -1)$ & $\frac{n}{R}$ \\
$W_{{\rm R}5}^{(n)+}(x, y)$ & $\bm{1}$ & $\bm{1}$ & $1$ 
& $0$ & $(-1, +1)$ & $\frac{n-\frac{1}{2}}{R}$ \\
$W_{{\rm R}5}^{(n)-}(x, y)$ & $\bm{1}$ & $\bm{1}$ & $-1$ 
& $0$ & $(-1, +1)$ & $\frac{n-\frac{1}{2}}{R}$ \\
\hline
$\varPhi_{\rm L}(x, y)$ & ${\bm{3}}$ & $\bm{2}$ & $0$ 
& $\frac{1}{6}$ & $(-1, -1)$ & $\frac{n}{R}$ \\
& $\bm{1}$ & $\bm{2}$ & $0$ 
& $-\frac{1}{2}$ & $(-1, +1)$ & $\frac{n-\frac{1}{2}}{R}$ \\
\hline
$\varPhi_{\rm R}(x, y)$ & $\overline{\bm{3}}$ & $\bm{1}$ & $\frac{1}{2}$
& $-\frac{1}{6}$ & $(-1, -1)$ & $\frac{n}{R}$ \\
& $\overline{\bm{3}}$ & $\bm{1}$ & $-\frac{1}{2}$ 
& $-\frac{1}{6}$ & $(+1, -1)$ & $\frac{n-\frac{1}{2}}{R}$ \\
& $\bm{1}$ & $\bm{1}$ & $\frac{1}{2}$ 
& $\frac{1}{2}$ & $(-1, +1)$ & $\frac{n-\frac{1}{2}}{R}$ \\
& $\bm{1}$ & $\bm{1}$ & $-\frac{1}{2}$ 
& $\frac{1}{2}$ & $(+1, +1)$ & $\frac{n}{R}$ \\
\hline
$\varPhi_{\rm B}(x, y)$ & $\bm{1}$ & $\bm{2}$ & $\frac{1}{2}$
& $0$ & $(+1, +1)$ & $\frac{n}{R}$ \\
& $\bm{1}$ & $\bm{2}$ & $-\frac{1}{2}$
& $0$ & $(+1, -1)$ & $\frac{n-\frac{1}{2}}{R}$ \\
\hline
\end{tabular}
\end{center}
\end{table}
\endgroup
In Table \ref{T-PSsp}, 
$Q_{\rm R}$ is the $U(1)_{\rm R}$ charge
and $Q_{\rm B-L}$ is the $U(1)_{\rm B-L}$ charge 
defined by
\begin{eqnarray}
Q_{\rm B-L} \equiv \sqrt{\frac{2}{3}} T_{\rm C}^{15},
\label{QB-L}
\end{eqnarray}
using the 15-th components of $T_{\rm C}^a$.
The fifth components of gauge bosons are would-be 
Nambu-Goldstone bosons and absorbed
by the corresponding 4D gauge bosons.

After the dimensional reduction,
we obtain the Lagrangian density:
\begin{eqnarray}
&~& \mathscr{L}_{\rm 4D}
= -\frac{1}{4} \sum_{a=1}^{8} G_{\mu\nu}^a G^{a \mu\nu}
-\frac{1}{4} \sum_{a=1}^{3} W_{\mu\nu}^a W^{a \mu\nu}
-\frac{1}{4} R_{\mu\nu} R^{\mu\nu}
-\frac{1}{4} N_{\mu\nu} N^{\mu\nu} 
\nonumber \\
&~&~~~~~~~~~ + (D_\mu \phi_{\rm R})^{\dagger} (D^\mu \phi_{\rm R}) 
+ (D_\mu \phi)^{\dagger} (D^\mu \phi) - V_{\rm 4D}
+ \mathscr{L}_{\rm KK},
\label{L-4D}
\end{eqnarray}
where $G_{\mu\nu}^a$, $W_{\mu\nu}^a$, $R_{\mu\nu}$
and $N_{\mu\nu}$ are field strengths of $SU(3)_{\rm C}$, $SU(2)_{\rm L}$,
$U(1)_{\rm R}$ and $U(1)_{\rm B-L}$ gauge bosons,
and $\mathscr{L}_{\rm KK}$ is the Lagrangian density
containing Kaluza-Klein modes.
Here, the covariant derivative $D_{\mu}$ 
and the scalar potential $V_{\rm 4D}$
are given by
\begin{eqnarray}
\hspace{-0.7cm}
&~& D_{\mu} = \partial_{\mu} + i g_3 \sum_{a=1}^{8} G_{\mu}^a T_{\rm C}^a
+ i g \sum_{a=1}^{3} W_{\mu}^a T_{\rm L}^a
+ i g_{\rm R} R_{\mu} Q_{\rm R} + i g_{\rm B-L} N_{\mu} Q_{\rm B-L},
\label{D_mu}\\
\hspace{-0.7cm}
&~& V_{\rm 4D}=\lambda_{\rm r} |\phi_{\rm R}|^4
+\lambda |\phi|^4 + \lambda_{\rm m} |\phi_{\rm R}|^2  |\phi|^2,
\label{V-4D}
\end{eqnarray}
respectively.
From the matching conditions between $\mathscr{L}_{\rm 5D}$
and $\mathscr{L}_{\rm 4D}$ at a scale $M_{\rm PS}$
above the compactification scale $M_{\rm c} (= 1/R)$,
we obtain the relations:
\begin{eqnarray}
&~& \left. g_3 = \sqrt{\frac{2}{3}} g_{\rm B-L} = g_4\right|_{M_{\rm PS}},~~ 
\left. g = g_{\rm L} = g_{\rm R}\right|_{M_{\rm PS}},~~
\label{cond-4D1}\\
&~& \left. \lambda_{\rm r} = \lambda_{\rm R}\right|_{M_{\rm PS}},~~
\left. \lambda = \lambda_{\rm B1}+ \lambda_{\rm B2}\right|_{M_{\rm PS}},~~
\left. \lambda_{\rm m} = \lambda_{\rm RB}\right|_{M_{\rm PS}}.
\label{cond-4D2}
\end{eqnarray}
Note that fields from zero modes are massless at $M_{\rm PS}$
and the value of $\lambda_{\rm r}$ does not necessarily agree with
that of $\lambda$ there.

\section{$SU(3)_{\rm C} \times SU(2)_{\rm L} \times U(1)_{\rm R}
\times U(1)_{\rm B-L}$ model}
\label{3211model}

Let us study 4D model with the gauge group
$SU(3)_{\rm C} \times SU(2)_{\rm L} \times U(1)_{\rm R}
\times U(1)_{\rm B-L}$ described by (\ref{L-4D}).
We refer to it as 3211 model.
Particle contents of massless fields
are listed in Table \ref{T-3211}.
\begingroup
\renewcommand{\arraystretch}{1.2}
\begin{table}[htbp]
\caption{Gauge quantum numbers
of massless fields in 4D 3211 model.}
\label{T-3211}
\begin{center}
\begin{tabular}{c|c|c|c|c|c|c}
\hline                              
particles & $SU(3)_{\rm C}$ & $SU(2)_{\rm L}$ & $Q_{\rm R}$ 
& $Q_{\rm B-L}$ & $Y$ & $Y_{\perp}$ \\ 
\hline\hline 
$G_{\mu}$ & $\bm{8}$ & $\bm{1}$ & $0$ & $0$ & $0$ & $0$\\
$W_{\mu}$ & $\bm{1}$ & $\bm{3}$ & $0$ & $0$ & $0$ & $0$\\
$R_{\mu}$ & $\bm{1}$ & $\bm{1}$ & $0$ & $0$ & $0$ & $0$\\
$N_{\mu}$ & $\bm{1}$ & $\bm{1}$ & $0$ & $0$ & $0$ & $0$\\
\hline
$\phi_{\rm R}$ & $\bm{1}$ & $\bm{1}$ & $-\frac{1}{2}$ 
& $\frac{1}{2}$ & $0$ & $\frac{5}{2}$\\
$\phi$ & $\bm{1}$ & $\bm{2}$ & $\frac{1}{2}$ & $0$ 
& $\frac{1}{2}$ & $-1$\\
\hline
$q_{{\rm L}A}$ & $\bm{3}$ & $\bm{2}$ & $0$ & $\frac{1}{6}$ 
& $\frac{1}{6}$ & $\frac{1}{2}$\\
$u_{{\rm R}A}$ & $\bm{3}$ & $\bm{1}$ & $\frac{1}{2}$ 
& $\frac{1}{6}$ & $\frac{2}{3}$ & $-\frac{1}{2}$\\
$d_{{\rm R}A}$ & $\bm{3}$ & $\bm{1}$ & $-\frac{1}{2}$ 
& $\frac{1}{6}$ & $-\frac{1}{3}$ & $\frac{3}{2}$\\
$l_{{\rm L}A}$ & $\bm{1}$ & $\bm{2}$ & $0$ & $-\frac{1}{2}$ 
& $-\frac{1}{2}$  & $-\frac{3}{2}$\\
$\nu_{{\rm R}A}$ & $\bm{1}$ & $\bm{1}$ & $\frac{1}{2}$ 
& $-\frac{1}{2}$ & $0$  & $-\frac{5}{2}$\\
$e_{{\rm R}A}$ & $\bm{1}$ & $\bm{1}$ & $-\frac{1}{2}$ 
& $-\frac{1}{2}$ & $-1$  & $-\frac{1}{2}$\\
\hline
\end{tabular}
\end{center}
\end{table}
\endgroup
In Table \ref{T-3211}, 
the subscript $A$ represents the generation of matter fields
on the 4D brane and runs from 1 to 3.
For a sake of reference, we denote values of the weak hypercharge
defined by $Y \equiv Q_{\rm R} + Q_{\rm B-L}$
and those of the $U(1)$ charge defined by 
$Y_{\perp} \equiv 5Q_{\rm B-L} - 2Y$,
which is orthogonal to $Y$.

\subsection{Running of gauge couplings}
\label{RGC-3211}

We study the running of gauge couplings.
By solving the renormalization group equations (RGEs)
of gauge couplings $g_i$
at the one-loop level,
we obtain the solutions,
\begin{eqnarray}
&~& \alpha_i^{-1}(\mu) = \alpha_i^{-1}(\mu_0)
- \frac{b_i}{2\pi} \ln\frac{\mu}{\mu_0}
\nonumber \\
&~& ~~~~~~~~~~~~
- \sum_{n=1} \frac{b'_i}{2\pi} \theta\left(\mu-\frac{n}{R}\right)
\ln\frac{\mu}{\frac{n}{R}}
- \sum_{n=1} \frac{b''_i}{2\pi} \theta\left(\mu-\frac{n-\frac{1}{2}}{R}\right)
\ln\frac{\mu}{\frac{n-\frac{1}{2}}{R}}
\label{g-RGE-sol}
\end{eqnarray}
where $\alpha_i \equiv g_i^2/(4\pi)$,
$\mu$ is a renormalization point,
$b_i$ are coefficients of $\beta$ functions for zero modes,
and $b'_i$ and $b''_i$ are coefficients of $\beta$ functions 
for Kaluza-Klein modes with masses $n/R$ and $(n-\frac{1}{2})/R$,
respectively.
The $\theta$ is a step function defined by
$\theta(x) = 1$ for $x > 0$, $\theta(x) = 0$ for $x < 0$
and $\theta(0) = 1/2$.
The values of $b_i$, $b'_i$ and $b''_i$ are listed in Table \ref{T-bi}.
\begingroup
\renewcommand{\arraystretch}{1.2}
\begin{table}[htbp]
\caption{Gauge couplings and their coefficients
of $\beta$ functions.}
\label{T-bi}
\begin{center}
\begin{tabular}{c|c|c|c|c|c}
\hline                             
& $SU(3)_{\rm C}$ & $SU(2)_{\rm L}$ & $U(1)_{\rm R}$ 
& $U(1)_{\rm B-L}$ & $U(1)_{\rm Y} $ \\
\hline\hline 
$g_i$ & $g_3$ & $g$ & $g_{\rm R}$ & $g_{\rm B-L}$ & $g_{\rm Y}$\\
\hline
$\alpha_i$ & $\alpha_3$ & $\alpha_2$ & $\alpha_{\rm R}$ 
& $\alpha_{\rm B-L}$ & $\alpha_{\rm Y}$\\
\hline
$b_i$ & $-7$ & $-\frac{19}{6}$ & $\frac{17}{4}$ 
& $\frac{11}{4}$ & $\frac{41}{6}$\\
\hline
$b'_i$ & $-1$ & $-\frac{1}{3}$ & $\frac{1}{2}$ 
& $\frac{1}{6}$ & --\\
\hline
$b''_i$ & $-\frac{1}{3}$ & $\frac{1}{3}$ & $-\frac{1}{2}$ 
& $-\frac{19}{18}$ & --\\
\hline
\end{tabular}
\end{center}
\end{table}
\endgroup
In Table \ref{T-bi}, we list $b_{\rm Y} = 41/6$ in the SM
for a sake of completeness, and -- represents not applicable.

By taking 
$M_{\rm PS} = n_{\varLambda}/R =n_{\varLambda} M_{\rm c}$
as $\mu$,
solutions are written by
\begin{eqnarray}
\hspace{-1cm}
&~& \alpha_i^{-1}(M_{\rm PS}) = \alpha_i^{-1}(\mu_0)
- \frac{b_i}{2\pi} \ln\frac{M_{\rm PS}}{\mu_0}
- \sum_{n=1}^{n_{\varLambda}} 
\frac{b'_i}{2\pi} \ln\frac{M_{\rm PS}}{\frac{n}{R}}
- \sum_{n=1}^{n_{\varLambda}} \frac{b''_i}{2\pi} 
\ln\frac{M_{\rm PS}}{\frac{n-\frac{1}{2}}{R}}
\nonumber \\
\hspace{-1cm}&~& ~~~~~~~~~~~~~ 
= \alpha_i^{-1}(\mu_0)
- \frac{b_i}{2\pi} \ln\frac{M_{\rm PS}}{\mu_0}
- \frac{b'_i}{2\pi} \ln\prod_{n=1}^{n_{\varLambda}}
\left(\frac{M_{\rm PS}}{nM_{\rm c}}\right)
- \frac{b''_i}{2\pi} \ln\prod_{n=1}^{n_{\varLambda}}
\left(\frac{M_{\rm PS}}{\left(n-\frac{1}{2}\right)M_{\rm c}}\right)
\nonumber \\
\hspace{-1cm}&~& ~~~~~~~~~~~~~ 
= \alpha_i^{-1}(\mu_0)
- \frac{b_i}{2\pi} \ln\frac{M_{\rm PS}}{\mu_0}
\nonumber \\
\hspace{-1cm}&~& ~~~~~~~~~~~~~~~~ 
- \frac{b'_i}{2\pi} \left(\frac{M_{\rm PS}}{M_{\rm c}}
\ln\frac{M_{\rm PS}}{M_{\rm c}}
- \ln\varGamma\left(\frac{M_{\rm PS}}{M_{\rm c}} + 1\right)\right)
\nonumber \\
\hspace{-1cm}&~& ~~~~~~~~~~~~~~~~ 
- \frac{b''_i}{2\pi} \left(\frac{M_{\rm PS}}{M_{\rm c}}
\ln\frac{M_{\rm PS}}{M_{\rm c}}
- \ln\varGamma\left(\frac{M_{\rm PS}}{M_{\rm c}} + \frac{1}{2}\right)
+ \ln\sqrt{\pi}\right),
\label{g-RGE-sol-MPS}
\end{eqnarray}
where $\varGamma$ is a gamma function defined by
\begin{eqnarray}
\varGamma(z) = \int_{0}^{\infty} t^{z-1} e^{-t} dt
\label{varGamma}
\end{eqnarray}
and we replace $\Pi_{n=1}^{n_{\varLambda}} n= n_{\varLambda} !$ 
and $\Pi_{n=1}^{n_{\varLambda}} \left(n - \frac{1}{2}\right)
= (2n_{\varLambda}-1)!!/2^{n_{\varLambda}}$
into $\varGamma(n_{\varLambda}+1)$ and 
$\varGamma\left(n_{\varLambda}+\frac{1}{2}\right)/\sqrt{\pi}$,
respectively.

\begin{figure}[ht!]
\begin{center}
\includegraphics[width=80mm]{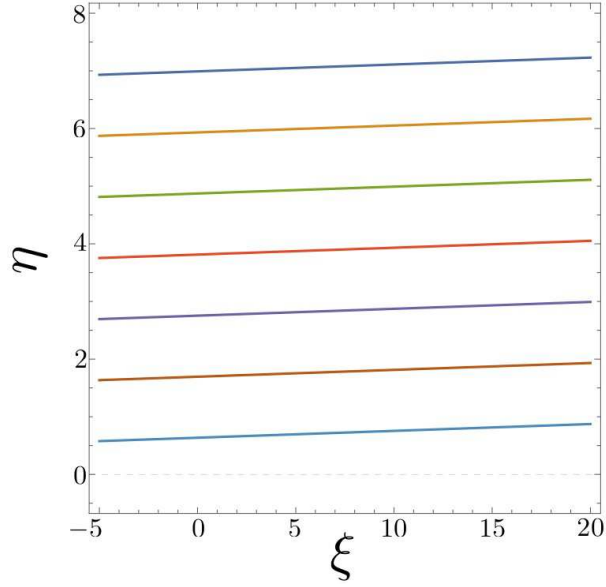}
\vskip-\lastskip
\caption{Allowed values of $\xi$ and $\eta$.
The colored lines represent the allowed values
for $M_{\rm c} =10^7$, $10^8$, $10^9$, $10^{10}$, $10^{11}$,
$10^{12}$ and $10^{13}$~GeV from the above.}
\label{Fig1}
\end{center}
\end{figure}

\begin{figure}[ht!]
\begin{center}
\includegraphics[width=130mm]{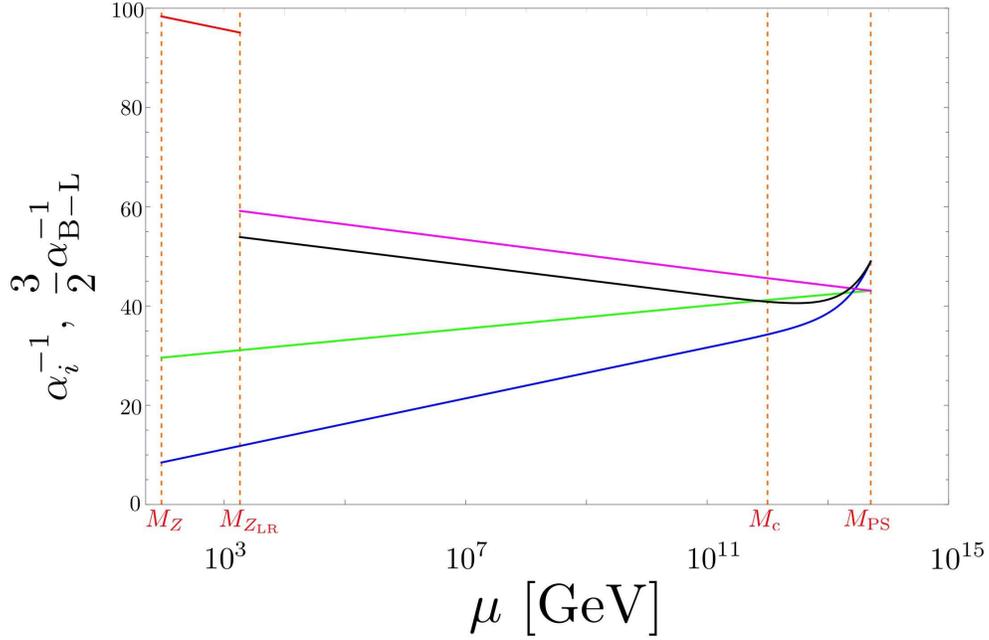}
\vskip-\lastskip
\caption{The running of gauge couplings.
The red, green, blue, violet and black lines stand for
the evolution of 
$\alpha_{\rm Y}^{-1}$, $\alpha_{2}^{-1}$,
$\alpha_{3}^{-1}$, $\alpha_{\rm R}^{-1}$
and $3 \alpha_{\rm B-L}^{-1}/2$, respectively.}
\label{Fig2}
\end{center}
\end{figure}

From the matching conditions at $M_{\rm PS}$ and $M_{Z_{\rm LR}}$,
we have the conditions:
\begin{eqnarray}
\left. \alpha_3 = \frac{2}{3} \alpha_{\rm B-L}\right|_{M_{\rm PS}},~~ 
\left. \alpha_2 = \alpha_{\rm R}\right|_{M_{\rm PS}},~~
\left. \alpha_{\rm Y}^{-1} = \alpha_{\rm R}^{-1} 
+ \alpha_{\rm B-L}^{-1}\right|_{M_{Z_{\rm LR}}},
\label{cond-gi}
\end{eqnarray}
where $M_{Z_{\rm LR}}$ is the mass of gauge boson
that becomes massive with the breakdown of
$U(1)_{\rm R} \times U(1)_{\rm B-L}$ into $U(1)_{\rm Y}$.
By combining with the solutions (\ref{g-RGE-sol-MPS}),
we obtain the sum rule:
\begin{eqnarray}
\hspace{-1cm}&~& {\alpha}_{\rm Y}^{-1}(M_Z) - \alpha_2^{-1}(M_Z)
 - \frac{2}{3} \alpha_3^{-1}(M_Z)
\nonumber \\
\hspace{-1cm}&~& ~~ = \frac{b_{\rm Y} - b_2 - \frac{2}{3} b_3}{2\pi}
\ln\frac{M_{\rm PS}}{M_Z}
+ \frac{-b_{\rm Y} + b_{\rm R} + b_{\rm B-L}}{2\pi}
\ln\frac{M_{\rm PS}}{M_{Z_{\rm LR}}},
\nonumber \\
\hspace{-1cm}&~& ~~~~ 
+ \frac{-b'_{\rm Y} - \frac{2}{3}b'_{3} + b'_{\rm R} + b'_{\rm B-L}}{2\pi}
\left(\frac{M_{\rm PS}}{M_{\rm c}}
\ln\frac{M_{\rm PS}}{M_{\rm c}}
- \ln\varGamma\left(\frac{M_{\rm PS}}{M_{\rm c}} + 1\right)\right)
\nonumber \\
\hspace{-1cm}&~& ~~~~ 
+ \frac{-b''_{\rm Y} - \frac{2}{3}b''_{3} + b''_{\rm R} + b''_{\rm B-L}}{2\pi}
\left(\frac{M_{\rm PS}}{M_{\rm c}}
\ln\frac{M_{\rm PS}}{M_{\rm c}}
- \ln\varGamma\left(\frac{M_{\rm PS}}{M_{\rm c}} + \frac{1}{2}\right)
+ \ln\sqrt{\pi}\right),
\label{gi-sumrule}
\end{eqnarray}
where $M_Z$ is the $Z$ boson mass 
given by $M_Z \doteqdot 91.19$GeV.
Using the values of ($b_i$, $b'_i$, $b''_i$) 
and the experimental values such that~\cite{PDG}
\begin{eqnarray}
\alpha_3^{-1}(M_Z) \doteqdot 8.467,~~
\alpha_2^{-1}(M_Z) \doteqdot 29.59,~~
\alpha_{\rm Y}^{-1}(M_Z) \doteqdot 98.36,
\label{alpha-exp}
\end{eqnarray}
we obtain the relation:
\begin{eqnarray}
M_{\rm PS} \doteqdot 3.675 \times 10^{13} \times (1.026)^{\xi}
\times \left(\frac{\sqrt{\pi} \varGamma(10^{\eta} + 1)}
{\varGamma(10^{\eta} + \frac{1}{2})}\right)^{0.1124}~{\rm GeV},
\label{MPS}
\end{eqnarray}
where $M_{Z_{\rm LR}}$ and $M_{\rm PS}$
are parametrized as $M_{Z_{\rm LR}} = 10^{\xi} \times M_Z$
and $M_{\rm PS} = 10^{\eta} \times M_{\rm c}$, respectively.
The factor including gamma functions represents
contributions from Kaluza-Klein modes.
From (\ref{MPS}), we find the interesting feature
that {\it the magnitude of $M_{\rm PS}$
is $O(10^{13})$~{\rm GeV} almost irrelevant to the value of 
$M_{Z_{\rm LR}}$}.
This is due to an accidental fact that
the coefficient of the second term in the right hand side 
of (\ref{gi-sumrule}) is tiny, i.e.,
$(-b_{\rm Y} + b_{\rm R} + b_{\rm B-L})/(2\pi) \doteqdot 0.02654$.
Further, the magnitude of $M_{\rm PS}$
is almost irrelevant to the value of $M_{\rm c}$,
because Kaluza-Klein modes appear as complete multiplets
(although there is a mass difference with $1/(2R)$)
with $(2/3) \times (b'_3 + b''_3) = b'_{\rm B-L} + b''_{\rm B-L} = -8/9$
and $b'_2 + b''_2 = b'_{\rm R} + b''_{\rm R} = 0$.\footnote{
It is pointed out that the running of gauge couplings
and the unification scale change drastically
due to the contributions from Kaluza-Klein modes
including incomplete multiplets~\cite{DD&G1, DD&G2}.
}
These features are understood from 
the $\xi$-$\eta$ plot satisfying (\ref{MPS}) given in Figure \ref{Fig1}.
\vspace{0mm}
Typical runnings of $\alpha_i^{-1}$ are depicted in Figure \ref{Fig2}.
\vspace{0mm}
Here we choose $\xi =1.3$, 
i.e., $M_{Z_{\rm LR}} \doteqdot 1819$~GeV,
and $M_{\rm c} = 1 \times 10^{12}$~GeV,
i.e., $\eta \doteqdot 1.7$, as a bench mark.\footnote{
The mass bound of an additional neutral gauge boson
of $SU(2)_{\rm L} \times SU(2)_{\rm R} \times U(1)$ (with $g = g_{\rm R}$)
is 630~GeV from $p\overline{p}$ direct search
and 1162~GeV from the electroweak fit~\cite{PDG}.
}


\subsection{Scalar potential in 3211 model}
\label{3211}

We study the breakdown of $U(1)_{\rm R} \times U(1)_{\rm B-L}$
and the electroweak symmetry.
The scalar potential at the tree level is given by $V_{\rm 4D}$ in (\ref{V-4D}).
The quartic couplings
$\lambda_{\rm r}$, $\lambda_{\rm m}$, $\lambda$ 
and the top Yukawa coupling $y_t$ obey the RGEs
at the one-loop level,
\begin{eqnarray}
&~& \frac{d \lambda_{\rm r}}{d \ln\mu}
= \frac{1}{16\pi^2} \biggl(20 \lambda_{\rm r}^2 + 2 \lambda_{\rm m}^2
- 3 g_{\rm R}^2 \lambda_{\rm r} - 3 g_{\rm B-L}^2 \lambda_{\rm r}
\nonumber \\
&~& ~~~~~~~~~~ \left.
+ \frac{3}{8} g_{\rm R}^4 + \frac{3}{4} g_{\rm R}^2 g_{\rm B-L}^2 
+ \frac{3}{8} g_{\rm B-L}^4\right),
\label{lambda-r-RGE}\\
&~& \frac{d \lambda_{\rm m}}{d \ln\mu}
= \frac{1}{16\pi^2} \left(4 \lambda_{\rm m}^2 
+ 8 \lambda_{\rm r} \lambda_{\rm m} + 12 \lambda \lambda_{\rm m}
- \frac{9}{2} g^2 \lambda_{\rm m} - 3 g_{\rm R}^2 \lambda_{\rm m} \right.
\nonumber \\
&~& ~~~~~~~~~~ \left. - \frac{3}{2} g_{\rm B-L}^2 \lambda_{\rm m} 
+ 6 y_t^2 \lambda_{\rm m} + \frac{3}{8} g_{\rm R}^4\right),
\label{lambda-m-RGE}\\
&~& \frac{d \lambda}{d \ln\mu}
= \frac{1}{16\pi^2} \left(24 \lambda^2 + \lambda_{\rm m}^2
- 3 g_{\rm R}^2 \lambda - 9 g^2 \lambda
+ \frac{3}{8} g_{\rm R}^4 + \frac{3}{4} g_{\rm R}^2 g^2 \right.
\nonumber \\
&~& ~~~~~~~~~~ \left.  + \frac{9}{8} g^4
+ 12 y_t^2 \lambda - 6 y_t^4\right),
\label{lambda-RGE-3211}\\
&~& \frac{d y_t}{d \ln\mu}
= \frac{1}{16\pi^2} \left(\frac{9}{2} y_t^3 - \frac{3}{4} g_{\rm R}^2 y_t
- \frac{1}{6} g_{\rm B-L}^2 y_t - \frac{9}{4} g^2 y_t - 8 g_3^2 y_t\right),
\label{yt-RGE-3211}
\end{eqnarray}
where the contributions from Kaluza-Klein modes
are omitted.

For a sake of completeness, we write down the RGEs of
the Higgs quartic coupling $\lambda$ and the top Yukawa coupling $y_t$
in the SM,
\begin{eqnarray}
&~& \frac{d \lambda}{d \ln\mu}
= \frac{1}{16\pi^2} 
\biggl(24 \lambda^2 - 3 g_{\rm Y}^2 \lambda - 9 g^2 \lambda
\nonumber \\
&~& ~~~~~~~~~ \left.+ \frac{3}{8} g_{\rm Y}^4 
+ \frac{3}{4} g_{\rm Y}^2 g^2 + \frac{9}{8} g^4
+ 12 y_t^2  \lambda - 6 y_t^4\right),
\label{lambda-RGE}\\
&~& \frac{d y_t}{d \ln\mu}
= \frac{1}{16\pi^2} \left(\frac{9}{2} y_t^3 - \frac{17}{12} g_{\rm Y}^2 y_t
- \frac{9}{4} g^2 y_t - 8 g_3^2 y_t\right).
\label{yt-RGE}
\end{eqnarray}
The $\lambda$ and $y_t$ run under the condition that
the SM ones match those of 3211 model at $M_{Z_{\rm LR}}$.

We obtain an effective potential improved by the RGEs
at the one-loop level,
\begin{eqnarray}
&~& V_{\rm eff}(\mu) =
\frac{\lambda_{\rm r}}{4} \varphi_{\rm R}^4
+ \frac{B_{\rm r}}{8} \varphi_{\rm R}^4 \left(\ln\frac{\varphi_{\rm R}^2}{\mu^2}
- \frac{25}{6}\right)
+ \frac{\lambda_{\rm m}}{4} \varphi^2 \varphi_{\rm R}^2
\nonumber\\
&~& ~~~~~~~~~ + \frac{B_{\rm m}}{4} \varphi^2 \varphi_{\rm R}^2 
\left(\ln\frac{\varphi\varphi_{\rm R}}{\mu^2} - 3\right)
+ \frac{\lambda}{4} \varphi^4
+ \frac{B}{8} \varphi^4 \left(\ln\frac{\varphi^2}{\mu^2}
- \frac{25}{6}\right),
\label{V-eff}
\end{eqnarray}
where $\varphi_{\rm R}^2 = 2\{({\rm Re}\phi_{\rm R})^2
+ ({\rm Im}\phi_{\rm R})^2\}$, 
$\varphi^2 = 2\{({\rm Re}\phi^+)^2
+ ({\rm Im}\phi^+)^2 + ({\rm Re}\phi^0)^2
+ ({\rm Im}\phi^0)^2\}$, $\varphi_{\rm R}^4 = (\varphi_{\rm R}^2)^2$,
$\varphi^4 = (\varphi^2)^2$, and $B_{\rm r}$, $B_{\rm m}$ and $B$
are given by,
\begin{eqnarray}
&~& B_{\rm r} = \frac{1}{16\pi^2} 
\left(20 \lambda_{\rm r}^2 + 2 \lambda_{\rm m}^2
+ \frac{3}{8} g_{\rm R}^4 + \frac{3}{4} g_{\rm R}^2 g_{\rm B-L}^2 
+ \frac{3}{8} g_{\rm B-L}^4\right),
\label{Br}\\
&~& B_{\rm m} = \frac{1}{16\pi^2} \left(4 \lambda_{\rm m}^2 
+ 8 \lambda_{\rm r} \lambda_{\rm m} + 12 \lambda \lambda_{\rm m}
+ \frac{3}{8} g_{\rm R}^4\right),
\label{Bm}\\
&~& B = \frac{1}{16\pi^2} \left(24 \lambda^2 + \lambda_{\rm m}^2
+ \frac{3}{8} g_{\rm R}^4 + \frac{3}{4} g_{\rm R}^2 g^2 + \frac{9}{8} g^4
- 6 y_t^4\right).
\label{B}
\end{eqnarray}

The effective potential $V_{\rm eff}(\mu)$
satisfies the renormalization conditions such that
\begin{eqnarray}
\left.\frac{\partial^4 V_{\rm eff}}{\partial \varphi_{\rm R}^4}
\right|_{\varphi_{\rm R}, \varphi = \mu} = \lambda_{\rm r}(\mu),~~
\left.\frac{\partial^4 V_{\rm eff}}{\partial \varphi_{\rm R}^2\partial\varphi^2}
\right|_{\varphi_{\rm R}, \varphi = \mu} = \lambda_{\rm m}(\mu),~~
\left.\frac{\partial^4 V_{\rm eff}}{\partial \varphi^4}
\right|_{\varphi_{\rm R}, \varphi = \mu} = \lambda(\mu)
\label{R-cond}
\end{eqnarray}
and does not depend on $\mu$, that is,
\begin{eqnarray}
&~& \frac{d V_{\rm eff}(\mu)}{d \ln\mu} = 
\left(\frac{\partial}{\partial\ln\mu} 
+ \frac{d \lambda_{\rm r}}{d \ln\mu}\frac{\partial}{\partial\lambda_{\rm r}}
+ \frac{d \lambda_{\rm m}}{d \ln\mu}\frac{\partial}{\partial\lambda_{\rm m}}
+ \frac{d \lambda}{d \ln\mu}\frac{\partial}{\partial\lambda}\right.
\nonumber \\
&~& ~~~~~~~~~~~~~~
\left. + \frac{d \varphi_{\rm R}}{d \ln\mu}\frac{\partial}{\partial\varphi_{\rm R}}
+ \frac{d \varphi}{d \ln\mu}\frac{\partial}{\partial\varphi} 
\right)V_{\rm eff}(\mu) = 0.
\label{mu-indep}
\end{eqnarray}

The first derivative of $V_{\rm eff}$ by fields are given by
\begin{eqnarray}
&~& \frac{\partial V_{\rm eff}}{\partial \varphi_{\rm R}} 
= \left\{\left(\lambda_{\rm r} + B_{\rm r} \ln\frac{\varphi_{\rm R}}{\mu}
-\frac{11}{6} B_{\rm r}\right) \varphi_{\rm R}^2\right.
\nonumber \\
&~& ~~~~~~~~~~ \left. + \frac{1}{2} \left(\lambda_{\rm m} 
+ B_{\rm m} \ln\frac{\varphi\varphi_{\rm R}}{\mu^2}
-\frac{5}{2} B_{\rm m}\right) \varphi^2\right\} \varphi_{\rm R},
\label{dVeff-R}\\
&~& \frac{\partial V_{\rm eff}}{\partial \varphi} 
= \left\{\left(\lambda + B \ln\frac{\varphi}{\mu}
-\frac{11}{6} B\right) \varphi^2\right.
\nonumber \\
&~& ~~~~~~~~~~ \left. + \frac{1}{2} \left(\lambda_{\rm m} 
+ B_{\rm m} \ln\frac{\varphi\varphi_{\rm R}}{\mu^2}
-\frac{5}{2} B_{\rm m}\right) \varphi_{\rm R}^2\right\} \varphi.
\label{dVeff}
\end{eqnarray}
From the stationary conditions 
\begin{eqnarray}
\left\langle \frac{\partial V_{\rm eff}}{\partial \varphi_{\rm R}} \right\rangle
= 0,~~ 
\left\langle \frac{\partial V_{\rm eff}}{\partial \varphi} \right\rangle
= 0,
\label{<dVeff>}
\end{eqnarray}
we obtain the relations:
\begin{eqnarray}
\left. \tilde{\lambda}_{\rm r} \langle \varphi_{\rm R} \rangle^2
= \frac{1}{2} \tilde{\lambda}_{\rm m} \langle \varphi \rangle^2
\right|_{\langle \varphi_{\rm R} \rangle},~~
\left. \tilde{\lambda} \langle \varphi \rangle^2
= \frac{1}{2} \tilde{\lambda}_{\rm m} \langle \varphi_{\rm R} \rangle^2
\right|_{\langle \varphi_{\rm R} \rangle},
\label{rels}
\end{eqnarray}
and, by combining them, the relation:
\begin{eqnarray}
\left. \tilde{\lambda}_{\rm r} 
= \frac{1}{4}\frac{\tilde{\lambda}_{\rm m}^2}{\tilde{\lambda}}
\right|_{\langle \varphi_{\rm R} \rangle},
\label{rel}
\end{eqnarray}
where $\tilde{\lambda}_{\rm r}$, $\tilde{\lambda}_{\rm m}$
and $\tilde{\lambda}$ are defined by
\begin{eqnarray}
&~& \tilde{\lambda}_{\rm r}(\mu)
\equiv \lambda_{\rm r} + B_{\rm r} \ln\frac{\langle\varphi_{\rm R}\rangle}{\mu}
-\frac{11}{6} B_{\rm r},
\label{tilde-lambda-r}\\
&~& \tilde{\lambda}_{\rm m}(\mu)
\equiv \lambda_{\rm m} 
+ B_{\rm m} \ln\frac{\langle\varphi\rangle\langle\varphi_{\rm R}\rangle}{\mu^2}
-\frac{5}{2} B_{\rm m},
\label{tilde-lambda-m}\\
&~& \tilde{\lambda}(\mu)
\equiv \lambda + B \ln\frac{\langle\varphi\rangle}{\mu}
-\frac{11}{6} B
\label{tilde-lambda}
\end{eqnarray}
and $|_{\langle \varphi_{\rm R} \rangle}$ means the value at 
$\mu = \langle \varphi_{\rm R} \rangle$.
We find that
{\it the breakdown of residual gauge symmetries
occurs radiatively via the Coleman-Weinberg mechanism,
such that
the $U(1)_{\rm R} \times U(1)_{\rm B-L}$ symmetry
is broken down to $U(1)_{\rm Y}$ 
at the scale $v_{\rm R} \equiv \langle \varphi_{\rm R} \rangle$
that satisfies {\rm (\ref{rel})}
and the $SU(2)_{\rm L} \times U(1)_{\rm Y}$ symmetry
is broken down to $U(1)_{\rm EM}$ by $\langle \varphi \rangle$}.
The hierarchy between $\langle \varphi_{\rm R} \rangle$
and $\langle \varphi \rangle$ comes from
the difference of magnitude among couplings
$\tilde{\lambda}_{\rm r}$, $\tilde{\lambda}_{\rm m}$
and $\tilde{\lambda}$, as seen from (\ref{rels}).

After the breakdown of $U(1)_{\rm R} \times U(1)_{\rm B-L}$, 
a gauge boson $Z_{{\rm LR}\mu}(x)$
acquires the mass
\begin{eqnarray}
M_{Z_{\rm LR}} = \frac{1}{2} \sqrt{g_{\rm R}^2 + g_{\rm B-L}^2}~v_{\rm R}.
\label{massE}
\end{eqnarray}
The $Z_{{\rm LR}\mu}(x)$ and $B_{\mu}(x)$ (a gauge boson
relating to $U(1)_{\rm Y}$) are given as linear combinations such that
\begin{eqnarray}
&~& Z_{{\rm LR}\mu}(x) = R_{\mu}(x) \cos\theta_{\rm R} 
- N_{\mu}(x) \sin\theta_{\rm R},
\label{Emu}\\
&~& B_{\mu}(x) = R_{\mu}(x) \sin\theta_{\rm R} 
+ N_{\mu}(x) \cos\theta_{\rm R},
\label{Bmu}
\end{eqnarray}
where the mixing angle $\theta_{\rm R}$ is defined by
$\tan\theta_{\rm R} \equiv g_{\rm B-L}/g_{\rm R}$.

Using the stationary conditions,
we obtain the following formula for mass matrix elements,
\begin{eqnarray}
&~& \left.
\left\langle \frac{\partial^2 V_{\rm eff}}{\partial \varphi_{\rm R}^2} \right\rangle
\right|_{v_{\rm R}}
= \left. \left(2\tilde{\lambda}_{\rm r} + B_{\rm r} 
- \frac{\tilde{\lambda}_{\rm m}}{4\tilde{\lambda}} B_{\rm m}\right)
\right|_{v_{\rm R}} v_{\rm R}^2,
\label{<d2Veff-R>}\\
&~&  \left.
\left\langle \frac{\partial^2 V_{\rm eff}}{\partial \varphi_{\rm R}\partial\varphi}
\right\rangle
\right|_{v_{\rm R}}
= \left. \left\langle \frac{\partial^2 V_{\rm eff}}{\partial \varphi
\partial\varphi_{\rm R}}
\right\rangle
\right|_{v_{\rm R}}
= \left. \left(\tilde{\lambda}_{\rm m} + \frac{B_{\rm m}}{2}\right)
\sqrt{- \frac{\tilde{\lambda}_m}{2\tilde{\lambda}}}
\right|_{v_{\rm R}} v_{\rm R}^2,
\label{<d2Veff-Rvarphi>}\\
&~& \left.
\left\langle \frac{\partial^2 V_{\rm eff}}{\partial \varphi^2} \right\rangle
\right|_{v_{\rm R}}
= \left. \left(2\tilde{\lambda} + B
- \frac{\tilde{\lambda}}{\tilde{\lambda}_{\rm m}} B_{\rm m}\right) 
\langle \varphi \rangle^2\right|_{v_{\rm R}}
\nonumber \\
&~& ~~~~~~~~~~~~~~~~~
= \left. \left(-\tilde{\lambda}_{\rm m} 
- \frac{\tilde{\lambda}_{\rm m}}{2\tilde{\lambda}} B 
+ \frac{B_{\rm m}}{2}\right)\right|_{v_{\rm R}} v_{\rm R}^2,
\label{<d2Veff>}
\end{eqnarray}
where $|_{v_{\rm R}}$ means the values at 
$\langle \varphi_{\rm R} \rangle = v_{\rm R}$.

Here we choose $\xi =1.3$, 
i.e., $M_{Z_{\rm LR}} \doteqdot 1819$~GeV,
and $M_{\rm c} = 1 \times 10^{12}$~GeV, i.e., $\eta \doteqdot 1.7$, 
as a bench mark.
In this case,
$v_{\rm R}$ is estimated as
\begin{eqnarray}
v_{\rm R} = 
\left. \frac{2M_{Z_{\rm LR}}}{\sqrt{g_{\rm R}^2 + g_{\rm B-L}^2}}
\right|_{M_{Z_{\rm LR}}}
\doteqdot 4854~{\rm GeV}
\label{vR-value}
\end{eqnarray}
and the mass matrix elements of scalar fields
are estimated as
\begin{eqnarray}
\left.\left(
\begin{array}{cc}
\left\langle \frac{\partial^2 V_{\rm eff}}{\partial \varphi_{\rm R}^2} 
\right\rangle
& 
\left\langle \frac{\partial^2 V_{\rm eff}}{\partial 
\varphi_{\rm R}\partial\varphi}
\right\rangle \\
\left\langle \frac{\partial^2 V_{\rm eff}}{\partial 
\varphi\partial\varphi_{\rm R}}
\right\rangle
& 
\left\langle \frac{\partial^2 V_{\rm eff}}{\partial \varphi^2} 
\right\rangle
\end{array}
\right)\right|_{v_{\rm R}}
\doteqdot \left(
\begin{array}{cc}
18138 & -721\\
-721 & 15961
\end{array}
\right)~{\rm GeV}^2.
\label{massmatrix-value}
\end{eqnarray}
After diagonalizing the mass matrix,
the mass of $\varphi_{\rm R}$-dominated component
is evaluated as
\begin{eqnarray}
m_{\rm R} \doteqdot 135~{\rm GeV}.
\label{massvarphiR-value}
\end{eqnarray}
The third term in the right hand side of (\ref{V-eff}) or (\ref{V-4D}) 
and its radiative corrections (4-th term in the right hand side of (\ref{V-eff}))
are Higgs portal.
By replacing $\varphi_{\rm R}$ into its VEV,
we obtain the following squared mass of Higgs boson 
approximately as
\begin{eqnarray}
m^2 \approx \frac{1}{2} (\lambda_{\rm m} - 3 B_{\rm m}) v_{\rm R}^2.
\label{m2}
\end{eqnarray}
From a numerical analysis,
we obtain the negative squared mass
because of $\lambda_{\rm m} < 3 B_{\rm m}$.
It can be interpreted that
the Higgs mechanism occurs effectively.

\begin{figure}[ht!]
\begin{center}
\includegraphics[width=140mm]{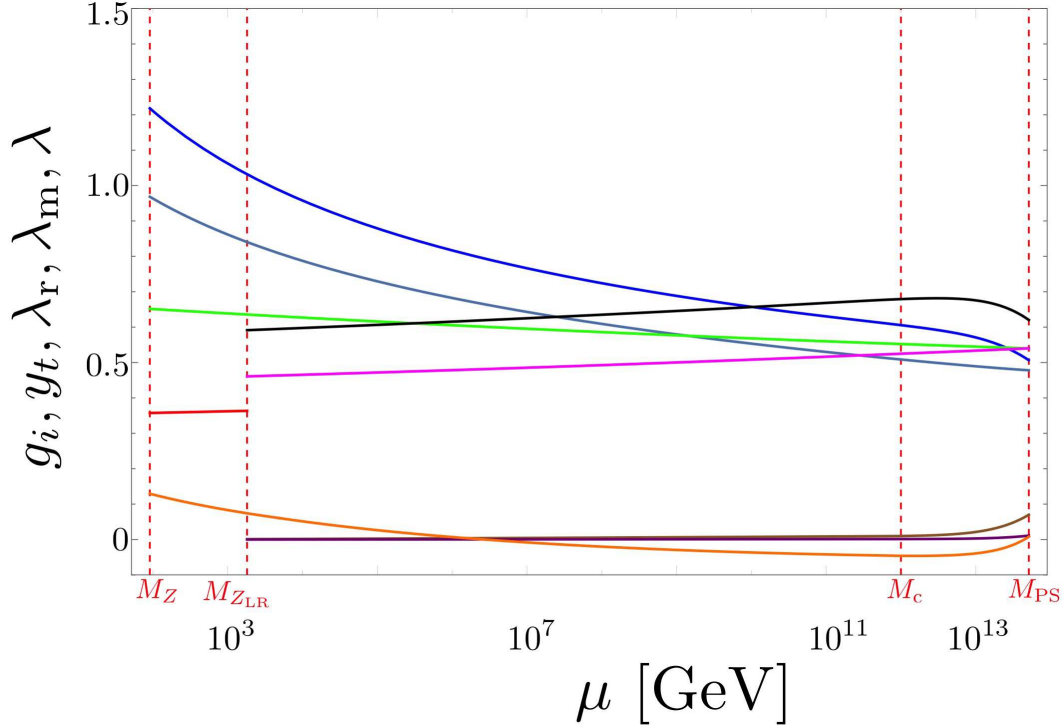}
\vskip-\lastskip
\caption{The running of various couplings.
The red, green, blue, violet, black, aqua, 
purple, dark brown and orange lines stand for
the evolution of 
$g_{\rm Y}$, $g_{2}$,
$g_{3}$, $g_{\rm R}$,
$g_{\rm B-L}$, $y_t$,
$\lambda_{\rm r}$, $\lambda_{\rm m}$ and
$\lambda$, respectively.}
\label{Fig3}
\end{center}
\end{figure}

The runnings of various couplings including
$\lambda_{\rm r}$, $\lambda_{\rm m}$ and
$\lambda$ are depicted in Figure \ref{Fig3}.
\vspace{0mm}
The values of $\lambda_{\rm r}$ and $\lambda_{\rm m}$ at $v_{\rm R}$
are estimated using stationary conditions (\ref{<dVeff>})
and $\lambda(v_{\rm R})$ with
$\langle \varphi \rangle \approx 246$~GeV.
Here, contributions from Kaluza-Klein modes
of gauge bosons are added, but those from 
Kaluza-Klein modes of scalar fields are not considered
because they are negligible small
when $\lambda_{\rm r}$, $\lambda_{\rm m}$ and
$\lambda$ take tiny values.
The running of $\lambda$ is almost same as that in the SM
because contributions from gluon and top quark are dominant.
From Figure 3, we find that the vacuum stability 
is recovered by the rapid increase of $\lambda$
due to contributions from Kaluza-Klein modes of gauge bosons.
We suppose that the vacuum stability problem
could be solved by changing the running
of $\lambda$ if $M_{\rm c}$ is less than $10^7$~GeV.
But, in this case, $\lambda$ can generally blow up infinity
much less than $M_{\rm PS}$ due to the threshold
corrections of various Kaluza-Klein modes.

\section{Conclusions and discussions}
\label{CD}

We have studied the origin of electroweak symmetry
under the assumption that 
$SU(4)_{\rm C} \times SU(2)_{\rm L} \times SU(2)_{\rm R}$
is realized on the 5D space-time $M^4 \times S^1/Z_2$.
The Pati-Salam type gauge symmetry is reduced to
$SU(3)_{\rm C} \times SU(2)_{\rm L} \times U(1)_{\rm R}
\times U(1)_{\rm B-L}$ at a high-energy scale $M_{\rm PS}$
above the compactification scale $M_{\rm c}$
by orbifold breaking mechanism on $S^1/Z_2$.
The breakdown of residual gauge symmetries
occurs radiatively via the Coleman-Weinberg mechanism,
such that the $U(1)_{\rm R} \times U(1)_{\rm B-L}$ symmetry
is broken down to $U(1)_{\rm Y}$ by the VEV
of an $SU(2)_{\rm L}$ singlet scalar field
and the $SU(2)_{\rm L} \times U(1)_{\rm Y}$ symmetry is broken down to
the electric one $U(1)_{\rm EM}$ by the VEV of the Higgs doublet,
using the negative squared mass originated from 
an interaction between the Higgs doublet 
and an $SU(2)_{\rm L}$ singlet scalar field as a Higgs portal.
The vacuum stability can be recovered
by the contributions from Kaluza-Klein modes
appearing at $M_{\rm c}$ and above there.

Our 3211 model has an excellent feature 
that $M_{\rm PS}$ is almost determined
as $M_{\rm PS} = O(10^{13})$ GeV from the gauge coupling unification
of $SU(3)_{\rm C}$ and $U(1)_{\rm B-L}$ into $SU(4)_{\rm C}$
and the left-right symmetry between 
$SU(2)_{\rm L}$ and $SU(2)_{\rm R}$.
On the contrary, the breaking scale $v_{\rm R}$
of $U(1)_{\rm R} \times U(1)_{\rm B-L}$
is not fixed from the information of gauge couplings alone.
The criterion of naturalness can favor $v_{\rm R}$ 
close to the weak scale. 

Our 3211 model has almost same particle contents
as those in a minimal $B-L$ extension of the SM
proposed in \cite{SK,IOO,IOO2,IO}.
Main differences of our model and the $B-L$ extended SM
are $U_{\rm B-L}$ charge assignment
of $SU(2)_{\rm L}$ singlet scalar field $\phi_{\rm R}$
and the interactions between
$U(1)$ gauge bosons and matter fields.
In our model, the $\nu_{{\rm R}A}$ and $\phi_{\rm R}$ have
$U(1)_{\rm B-L}$ charge of $-1/2$ and $1/2$, respectively.
Then, allowed interaction terms between them
are not renormalizable ones but non-renormalizable ones,
e.g., $(f_{AB}/\Lambda)\phi_{\rm R}^2 
\overline{\nu}^c_{{\rm R}A}\nu_{{\rm R}A}$,
where $\Lambda$ is a high-energy scale such as $M_{\rm PS}$.
Hence small Majorana masses appear after the breakdown of
$U(1)_{\rm R} \times U(1)_{\rm B-L}$ and the seesaw mechanism 
does not work at the TeV scale.
In this paper, we focus on physics of gauge symmetry breaking sector.
It would be meaningful to investigate flavor physics 
relating to quarks and leptons in our model.
It would be also important
to clarify the relationship between our model and the $B-L$ extended SM
through the study of gauge kinetic mixing and so on.

\section*{Acknowledgments}
The authors acknowledge Yasunari Nishikawa
for collaborations in the early stages of this work.
The authors thank Prof. S. Iso for valuable discussions.
This work was supported in part by scientific grants 
from Iwanami Fu-Jukai and the MEXT-Supported Program 
for the Strategic Research Foundation at Private Universities 
“Topological Science” under Grant No.~S1511006 (Y.~G.)
and from the Ministry of Education, Culture,
Sports, Science and Technology under Grant No.~17K05413 (Y.~K.).

\end{document}